\documentclass{aastex}
\usepackage{natbib}
\usepackage{verbatim}
\usepackage{epsfig}
\usepackage{emulateapj5}

\newcommand{\fdot}{$\dot{f}$}
\newcommand{\ffdot}{$f$--$\dot{f}$}
\newcommand{\psr}{PSR~J1807$-$2459}
\newcommand{\sm}{\ifmmode\mbox{M}_{\odot}\else$\mbox{M}_{\odot}$\fi}
\newcommand{\degrees}{\ifmmode^{\circ}\else$^{\circ}$\fi}
\newcommand{\amin}{\ifmmode^{\prime}\else$^{\prime}$\fi}
\newcommand{\asec}{\ifmmode^{\prime\prime}\else$^{\prime\prime}$\fi}
\newcommand{\simgt}{\lower.5ex\hbox{$\; \buildrel > \over \sim \;$}}
\newcommand{\simlt}{\lower.5ex\hbox{$\; \buildrel < \over \sim \;$}}

\slugcomment{Accepted by ApJ Letters on 2000 October 11}

\shortauthors{Ransom et al.}
\shorttitle{Binary Millisecond Pulsar in NGC6544}

\begin{document}

\title{A Binary Millisecond Pulsar in Globular Cluster NGC6544}

\author{Scott~M.~Ransom\altaffilmark{1}, 
  Lincoln~J.~Greenhill\altaffilmark{1}, 
  James~R.~Herrnstein\altaffilmark{2, 3}, 
  Richard~N.~Manchester\altaffilmark{4}, 
  Fernando~Camilo\altaffilmark{5}, 
  Stephen~S.~Eikenberry\altaffilmark{6}, 
  Andrew~G.~Lyne\altaffilmark{7}}


\altaffiltext{1}{Harvard-Smithsonian Center for Astrophysics, 
  60 Garden St., Cambridge, MA 02138; 
  ransom@cfa.harvard.edu, lgreenhill@cfa.harvard.edu}
\altaffiltext{2}{National Radio Astronomy Observatory, 
  P.O. Box O, Socorro, NM 87801}
\altaffiltext{3}{Current address: Renaissance Technologies
  Corporation, 600 Route 25A, East Setauket, NY 11733-1249; 
  jrh@rentec.com}
\altaffiltext{4}{Australia Telescope National Facility, CSIRO, 
  P.O. Box 76, Epping, NSW 1710, Australia; 
  rmanches@atnf.csiro.au}
\altaffiltext{5}{Columbia Astrophysics Laboratory, Columbia University, 
  550 West 120th St., New York, NY 10027; 
  fernando@astro.columbia.edu}
\altaffiltext{6}{Astronomy Department, 212 Space Sciences 
  Building, Cornell University, Ithaca, NY 14853; 
  eiken@astrosun.tn.cornell.edu}
\altaffiltext{7}{University of Manchester, Jodrell Bank Observatory, 
  Macclesfield, Cheshire, SK11 9DL, UK; 
  agl@jb.man.ac.uk}

\begin{abstract}
  We report the detection of a new 3.06\,ms binary pulsar in the
  globular cluster NGC6544 using a Fourier-domain ``acceleration''
  search.  With an implied companion mass of $\sim0.01\,\sm$ and an
  orbital period of only $P_b\sim1.7$\,hours, it displays very similar
  orbital properties to many pulsars which are eclipsed by their
  companion winds.  The orbital period is the second shortest of known
  binary pulsars after 47~Tuc~R.  The measured flux density of
  $1.3\pm0.4$\,mJy at 1332\,MHz indicates that the pulsar is almost
  certainly the known steep-spectrum point source near the core of
  NGC6544.
\end{abstract}


\section{Introduction}
\label{sec:intro}
Globular clusters are rich sources of radio pulsars.  Since the
discovery of the first pulsar in a cluster by \citet{lbm+87}, a series
of deep searches have accumulated almost 50 pulsars in globular
clusters, with the majority of these being binary millisecond pulsars
(MSPs).  See \citet{lyn95} and \citet{ka96} for reviews.

Fruchter \& Goss (1990; 2000, hereafter FG00) \nocite{fg90, fg00} used
the Very Large Array (VLA) to detect steep-spectrum radio sources,
which are likely pulsars, in numerous globular clusters.  They
discovered a relatively bright point-like source within $5''$ of the
optical center of cluster NGC6544 ($1.2\pm0.1$\,mJy at 20\,cm and
$6.5\pm0.8$\,mJy at 90\,cm).  \citet{bl96} searched NGC6544 at 50\,cm
as part of a survey of globular clusters and found no pulsars down to
a limiting flux density of $\sim4$\,mJy.

It has been argued that Doppler smearing from orbital motion could
explain the lack of detection of pulsations from the pulsar-like
sources found in VLA surveys (FG00).  The sensitivities of
observations whose durations exceed even a few percent of the orbital
period may be drastically reduced for traditional analyses
\citep{jk91}.  Various schemes have been developed to correct, at
least partially, for this loss of sensitivity when the Doppler effect
of the orbital motion can be approximated as a constant frequency
derivative \citep[e.g.][]{mk84, agk+90, whn+91}.  These
``acceleration'' searches have met with significant success, including
the discovery of nine new binary MSPs in the globular cluster 47~Tucanae
\citep{clf+00}.

We have discovered a binary MSP in NGC6544 using Fourier-domain
techniques to correct for pulsar orbital motion.  This pulsar is
almost certainly the radio source reported by FG00 near the center of
the cluster.  Results from our single observation indicate that this
object shares many of the properties displayed by the burgeoning class
of eclipsing binary MSPs.  These systems are characterised by short
orbital periods ($P_b \sim 1-10$\,hours) and very low-mass companions
($m_2 \sin i \simlt 0.05\,\sm$)\ \citep{rpr00, nzt00}.  \citet{dlm+00}
independently discovered this pulsar and report the results of
follow-up observations.

\section{Observations and Data Reduction}
\label{sec:data}

We observed NGC6544, a dense core-collapsed globular cluster, on 1998
February~7 with the multibeam data acquisition system on the Parkes
radio telescope in Australia.  Signals from each of two orthogonal
linear polarizations were measured and summed from 256 contiguous
0.25\,MHz wide channels centered at 1332\,MHz.  Each
channel was one-bit sampled at 8\,kHz and written to magnetic tape.
The full observation comprised 28.9 minutes of data.  \citet{swb+96}
and \citet{lcm+00} discuss the observing system in detail.

Since no pulsars were known in this cluster before our observation, we
searched a wide range of possible dispersion measures (DMs) for
candidate signals.  We dedispersed the data into time series of
13865600 points using 600 trial DMs from 0 to 600\,pc\,cm$^{-3}$, in
increments of 1\,pc\,cm$^{-3}$.  This stepsize allowed a worst-case
dispersion smearing over our bandwidth of $\sim0.1$\,ms.  After
barycentering the data using the JPL~DE200 ephemeris \citep{sta82}, we
performed a Fast Fourier Transform (FFT) on each time series and saved
the resultant spectra to disk.

Traditional time-domain acceleration searches are performed by
stretching or compressing a time series to compensate for a constant
frequency derivative and then Fourier transforming the resulting
series.  We have developed a Fourier-domain acceleration search
(S.~Ransom \& S. Eikenberry 2000, in preparation) to compute the
coherent Fourier response over portions of the frequency--frequency
derivative (\ffdot) plane using only local Fourier amplitudes from an
FFT of the whole dataset.  Similar techniques were described in
\citet{mdk93}.

The method correlates a predicted Fourier response with a subset of
the complex Fourier amplitudes in the initial full-length FFT.  We
effectively apply a matched filter to the data which ``sweeps up'' all
of the signal power that the orbital motion spreads over nearby
frequency bins.

Fourier-domain acceleration searches offer several significant
advantages over their time-domain counterparts:
\begin{enumerate}
\item Fourier methods do not require stretching or compressing the
  time series.  Time-domain stretching or compressing is usually
  performed using linear interpolation, which changes the statistics
  of the data.  Portions of the data are effectively two-bin averaged
  which decreases sensitivity to high frequency signals.
\item Time-domain techniques require a full-length FFT for each trial
  acceleration.  For very long integrations (which are common in
  cluster searches) that do not fit into the computer's core memory,
  FFTs take orders-of-magnitude more time to compute than usual
  \citep[e.g.][]{bai90a}.  The Fourier technique requires only a
  single full-length FFT for each observation (or trial DM).
\item Correlations using only localized Fourier amplitudes are fast
  and memory efficient.  The correlations can always be performed in
  core memory and are calculated efficiently using short FFTs and
  pre-computed response templates.  The memory locality also allows
  efficient parallelization of Fourier-domain searches.
\item The Fourier-domain method allows the calculatation of only
  independent \ffdot\ trials.  With time-domain techniques, as
  described by \citet{clf+00}, any choice of acceleration stepsize
  results in either under- or over-sampling accelerations for the vast
  majority of frequencies searched.  This amounts to loss of
  sensitivity or wasted CPU cycles respectively.  A small amount of
  over-sampling can be incorporated to eliminate ``scalloping'' in
  both the $f$ and \fdot\ directions (see e.g. \citealt{van89}).
\end{enumerate}

We performed Fourier-domain acceleration searches on each of the long
FFTs looking for signals that drifted by up to 100 Fourier frequency
bins during the observation.  If we define $z$ to be the number of
Fourier bins that the pulsar frequency ($f_o$) drifts over the course
of an integration, the corresponding average acceleration of the
pulsar is $a = zc/(T^2f_o)$, where $T$ is the total integration time.
For this observation, a frequency drift of 100 Fourier bins corresponds
to an acceleration of 500\,m\,s$^{-2}$ for a 10\,Hz signal or
5\,m\,s$^{-2}$ for a 1000\,Hz signal.  Since most known binary pulsars
show maximal accelerations of only a few m\,s$^{-2}$, we were
sensitive to all but the most exotic binaries or pulsars with spin
periods ($P_{spin}$) much less than $\sim2$\,ms.  Pulse profiles of
strong candidates were folded at a series of trial DMs, frequencies,
and frequency derivatives around the acceleration search values in
order to maximize the signal-to-noise.

\bigskip
\centerline{\psfig{figure=fig1color.eps,
    height=3.25in,angle=0,clip=}} 
\noindent {\footnotesize {\bf Figure 1} ---
  An $18\,\sigma$ (single trial) detection of the 1.7\,hour binary
  \psr\ in the globular cluster NGC6544 using a Fourier-domain
  ``acceleration'' search.  Contour intervals correspond to 30, 60,
  90, 120, and 150 times the average local power level.  The intrinsic
  pulsar period and $\dot{f} = 0$ (which corresponds to an
  un-accelerated FFT of the data) are marked by the solid black lines.
  The dashed line is the predicted ``path'' of the pulsar given the
  orbital solution in Table~1.  During the 28.9\,minute
  observation, the pulsar moved from $\sim11$ o'clock to $\sim3$
  o'clock on the ellipse.  The peak's slight offset from the ellipse
  and the presence of ``shoulders'' indicate that the constant \fdot\ 
  assumption of the acceleration search could not fully correct for
  the orbital motion during this observation.}
\bigskip

\section{Results}
\label{sec:results}

We detected a strong pulsar candidate (power/average local power
$\sim165$ corresponding to a single trial significance of
$18\,\sigma$) at a DM of 134\,pc\,cm$^{-3}$.  The average frequency
was 326.85895\,Hz and the signal had drifted by 8.6 Fourier bins
($a=2.6$\,m\,s$^{-2}$) during the observation (see Figure 1).  Search
techniques that do not account for acceleration would have detected
the signal at about $\sim34$ times the average local power level,
corresponding to an $8\,\sigma$ detection.

Optimization of the candidate by pulse-folding produced a narrow
pulsar-like profile at a localized maximum in DM of 134\,pc\,cm$^{-3}$
indicating that the signal was almost certainly not due to terrestrial
interference.  The \citet{tc93} free-electron model gives an estimated
distance of 3.3\,kpc ($\sim25$\% error) for this DM in the direction
of NGC6544, in fair agreement with the published cluster distance of
$\sim2.6$\,kpc obtained by fitting the mean {\em V}\ magnitudes of
horizontal branch stars \citep{har96}.

We estimated a pulsar flux density at 1332\,MHz of $1.3\pm0.4$\,mJy by
comparing the integrated pulsed flux and observed noise to the
predicted system noise level given a system sensitivity and total
system temperature of $T_{tot} \sim T_{sys} + T_{sky}$.  The multibeam
system has a cold-sky $T_{sys} \sim 21\,$K and a sensitivity of
1.36\,Jy\,K$^{-1}$.  \citet{hss+82} measured a sky temperature at
408~MHz in the direction of NGC6544 of $T_{sky} \sim 260\,$K.
Assuming a spectral index of $-2.6$, which is typical for this region
\citep{lmop87}, we obtain $T_{sky} \sim 12\,$K and $T_{tot} \sim
33\,$K at 1332\,MHz.  FG00 reported a 20\,cm flux density of
$1.2\pm0.1$\,mJy and a spectral index of $-1.1$ between 90\,cm and
20\,cm which corresponds to a flux density of $\sim 1.4 \pm 0.1$\,mJy
at 1332\,MHz.  Our measured flux density agrees with that of FG00
within the errors, suggesting that the pulsar may be the previously
unidentified point source near the core of NGC6544.  Since the
DM-based distance estimate for the pulsar matches published values for
NGC6544, cluster membership is very likely.

The relatively large signal acceleration indicates that the pulsar is
in a short period binary with a low-mass companion.  These systems,
almost without exception, have eccentricities very near to zero due to
tidal circularization (see e.g. \citealt{pk94} and \citealt{rh95}).

\bigskip
\centerline{\psfig{figure=fig2.eps,
    height=3.25in,angle=0,clip=}} 
\noindent {\footnotesize {\bf Figure 2} ---
  Orbital fit for \psr\ assuming an eccentricity of zero.  The
  error-bars plotted are $\pm2\,\sigma$ uncertainties in the pulse
  arrival times.  An eclipse, if present, should occur near True
  Anomaly$\;= 90\degrees$, which corresponds to superior conjunction.
  The apparent lack of an eclipse could be due to the relatively high
  observing frequency (1332\,MHz).  The inset plot shows the observed
  pulse profile after folding with the best-fit orbital ephemeris and
  pulsar frequency.  The error-bar indicates $\pm2\,\sigma$
  uncertainties in the profile values.}
\centerline{\psfig{figure=fig3.eps,
    height=3.25in,angle=0,clip=}} 
\noindent {\footnotesize {\bf Figure 3} ---
  Error ellipses (from inner to outer of $1\,\sigma$, $2\,\sigma$, and
  $3\,\sigma$) for the orbital period and semi-major axis of \psr.
  The large covariance is due to the fact that the observation
  covered only $\sim30\%$ of the orbital period.}
\bigskip

In order to study the assumption that the pulsar is a member of a
low-mass binary with an eccentricity of zero, we split the observation
into 16 equal-length parts and folded the data in each at the nominal
pulsar frequency.  We determined phase offsets for each of the pulse
arrival times and performed a Levenberg-Marquardt least-squares fit of
a circular orbit (i.e. sinusoid) to the phase offsets.  The results
are tabulated in Table~1 and the residuals and error ellipses of the
fit are shown in Figures~2 and 3 respectively.

\bigskip
\begin{center}
\begin{tabular}{lc}
\multicolumn{2}{c}{\sc Table 1} \\
\multicolumn{2}{c}{Parameters for \psr} \\
\tableline
\tableline
Parameter & Value \\
\tableline
Right Ascension\tablenotemark{a} $\;$ (J2000) & 
18$^{\rm h}$ 07$^{\rm m}$ 20\fs36(2) \\
Declination\tablenotemark{a} $\;$ (J2000) & 
$-24\degrees\;59\amin\;52\farcs6(4)$ \\
Dispersion Measure (pc cm$^{-3}$) & 134(2) \\
Flux Density at 1332\,MHz (mJy) & 1.3(4) \\
Pulse FWHM at 1332\,MHz (\%) & 11(3) \\
Pulsar Period (s) & 0.003059447(2) \\
Epoch (MJD) & 50851.9 \\
Orbital Period (days) & 0.070(3) \\
Projected Semi-Major Axis (lt-s) & 0.0116(8) \\
Eccentricity\tablenotemark{b} & 0.0 \\
Epoch of Periastron (MJD) & 50851.887(1) \\
Mass Function (\sm) & $3.4(5)\times10^{-7}$ \\
Companion Mass Limit\tablenotemark{c} $\;$ (\sm) & $\geq 0.009$ \\
\tableline
\multicolumn{2}{l}{$^{a}\;$A.~Fruchter 2000, private communication.} \\
\multicolumn{2}{l}{\hspace{0.5cm}Fruchter \& Goss discovered an error in the 20\,cm} \\
\multicolumn{2}{l}{\hspace{0.5cm}position reported in FG00.  The corrected position} \\
\multicolumn{2}{l}{\hspace{0.5cm}shown above will appear in an erratum.} \\
\multicolumn{2}{l}{$^{b}\;$Parameter assumed to be exactly zero.
  See \S\ref{sec:results}.} \\
\multicolumn{2}{l}{$^{c}\;$Assuming a pulsar mass of $m_1=1.4\,\sm$.} \\
\multicolumn{2}{l}{{\sc Note} --- Numbers in parentheses 
  represent 2$\,\sigma$} \\
\multicolumn{2}{l}{\hspace{0.5cm}uncertainties in the last digit quoted.} \\
\tableline
\end{tabular}
\end{center}

We found no other candidates in our search of NGC6544 to a limiting
flux density of $\sim1.1$\,mJy for 1\,ms period signals and
$\sim0.5$\,mJy for periods $\simgt10$\,ms.


\section{Discussion}
\label{sec:discuss}

There are currently eight binary pulsar systems that display similar
orbital characteristics to those of \psr: PSR~B1957$+$20 \citep{fst88}
and PSR~J2051$-$0827 \citep{sbl+96} in the Galactic disk,
PSR~J1910$+$0004 in NGC6760 \citep{dma+93}, and pulsars 47~Tuc I, J,
O, P, and R \citep{clf+00}.  Each of these systems displays a pulse
period of a few milliseconds, an orbital period of 1-10\,hours, and a
very low-mass companion ($m_2 \sin i \simlt 0.05\,\sm$).  Some of
these systems display radio eclipses, particularly at longer
wavelengths (i.e. 50-90\,cm).

Our current 1332-MHz data show no evidence for an eclipse (see
Figure~4) at or near superior conjunction, but we cannot rule out the
possiblity of eclipses at longer wavelengths.  Pulsars J2051$-$0827
and 47~Tuc~J show eclipses at radio wavelengths between 50-90\,cm, but
are detected at all orbital phases at 20\,cm, scintillation permitting
\citep{sbl+96, clf+00}.  PSR~J1910$+$0004 in NGC6760 was detected at
all orbital phases at 20\,cm, but \citet{dma+93} could not rule out
short duration eclipses or eclipses at longer wavelengths.

Most formation theories for short-period binary MSPs in globular
clusters involve dynamical interactions with primordial binaries which
have been shown to exist in significant numbers \citep{hmg+92}.
\citet{rpr00} describe a recent variant of the standard ``spin-up''
model \citep[see][for a review]{pk94} using an exchange interaction
between a neutron star (NS) and a hard primordial binary containing at
least one relatively massive main sequence (MS) star ($\sim$
1-3\,\sm).  If the exchange interaction results in a binary containing
the NS and the massive MS star, the system enters a common-envelope
(CE) phase once the MS star evolves and fills its Roche lobe.  At the
end of the CE phase, the NS emerges with a very low-mass companion in
a very short period circular orbit.  Systems with orbital periods
$P_b\simlt8$\,hours undergo evolution due to gravitational radiation
and enter a second phase of mass transfer that spins the NS up to
millisecond periods.

The scenario described by \citet{rpr00} is possible only if the
exchange interactions occurred at a time when massive MS stars were
still present in the cluster.  The NSs and primordial binaries must
also have undergone mass segregation and ended up near the cluster
core before the MS stars evolved.  The timescale for mass segregation
is the half-mass relaxation time, which for NGC6544 is $\sim200$\,Myr
\citep{har96}.  This is significantly less than the MS lifetime of a
1-3\,\sm\ star and implies that this scenario may have produced \psr.

Alternatives to ``spin-up'' models exist in which the NS is created
either by accretion induced collapse of a white dwarf (WD)
\citep[e.g.][]{bg90}, or the coalescence of a pair of massive WDs
\citep{cl93}.  Pairs of primordial binaries interact to produce hard
binaries which contain a massive WD or WDs that these models require.
The NSs that result from these models are born spinning rapidly
($P_{spin} \sim 2-10$\,ms), have relatively weak ($10^{9}-10^{10}$\,G)
magnetic fields characteristic of MSPs, and may have low-mass
companions in very short period orbits.  Interestingly, these systems
could be the progenitors of the low-mass X-ray binaries (LMXBs) rather
than their descendants \citep{cmr93}.

Future observations of \psr\ will lead to improved measurements of
rotational and orbital parameters, an estimate of or upper limit to
the orbital eccentricity, and perhaps information about the
circumstellar medium if eclipses are detected.  \citet{rpr00} predict
that many similar systems must exist in the globular cluster system
--- some with orbital periods as short as $\sim15$\,min.  If recent
successes such as the 20\,cm searches of 47~Tucanae \citep{clf+00} are
any indication, this prediction will soon be put to the test.  The
ever-increasing speed of computers and improved algorithms for binary
pulsar detection \citep[e.g.][]{ran00} will allow the analysis of
extremely long observations (i.e. days or weeks) that should reveal
even the weakest of pulsars in the tightest of binaries.

\acknowledgements We would like to thank Vicky Kaspi, Josh Grindlay,
Giovanni Fazio, and Nichi D'Amico for useful discussions.  FC is
supported by NASA grant NAG~5-3229.  SE is supported by a NSF CAREER
grant.


\clearpage

\end{document}